\documentclass[preprint,showpacs,preprintnumbers,amsmath,amssymb,aps]{revtex4}

\usepackage{graphicx}
\usepackage{dcolumn}
\usepackage{bm}
\usepackage{epsfig}

\begin{document}


\title{Quantum dynamics of the $\mbox{Li}+\mbox{HF}\rightarrow\mbox{H}+\mbox{LiF}$
 reaction at ultralow temperatures}

\author{P. F. Weck}
\email{weckp@unlv.nevada.edu}
\author{N. Balakrishnan}
\email{naduvala@unlv.nevada.edu}

\affiliation{Department of Chemistry, University of Nevada Las Vegas, 
 4505 Maryland Parkway, Las Vegas, NV 89154, USA}

\date{\today}

\begin{abstract}

Quantum mechanical calculations are reported for the  
$\mbox{Li}+\mbox{HF}(v=0,1,j=0)\rightarrow\mbox{H}+\mbox{LiF}(v',j')$ 
bimolecular scattering process at low and ultralow temperatures. Calculations 
have been performed for zero total angular momentum 
using a recent high accuracy potential energy surface 
for the $X^2A'$ electronic ground state. For 
$\mbox{Li}+\mbox{HF}(v=0,j=0)$, the reaction is dominated by resonances 
due to the decay of metastable states of 
the $\mbox{Li}\cdots\mbox{F}-\mbox{H}$ van der Waals complex.     
Assignment of these resonances has been carried out by calculating the eigenenergies 
of the quasibound states. 
We also find that while chemical reactivity is greatly enhanced by 
vibrational excitation the resonances get mostly washed out in the reaction 
of vibrationally excited HF with Li atoms. 
In addition, we find that at low energies, 
the reaction is significantly suppressed due to the formation of rather deeply 
bound van der Waals complexes and the less efficient tunneling of the relatively 
heavy fluorine atom.    
 
\end{abstract}

\pacs{33.70.-w}
\maketitle

\section{Introduction \label{sec:intro}} 

The past few years have witnessed an extremely prolific research 
effort in the experimental and theoretical investigation of 
ultracold molecules. The rapid development of techniques for cooling, 
trapping, and manipulating molecules at ultracold 
temperatures \cite{bet00,joc03,gre03,zwi03,cub03} led recently to the 
creation of Bose-Einstein condensates (BEC) of diatomic 
molecules \cite{joc03,gre03,zwi03}. This major achievement opens new perspectives 
in the exploration of the crossover regime between BEC and Bardeen-Cooper-Schrieffer 
(BCS) superfluidity \cite{tim01,reg04,bar04,bou04}, as well as in the conception of 
qubits in quantum computers using electric dipole moment couplings between ultracold 
polar molecules \cite{bar95,bre99,pla99,dem02}. 

Among the wealth of techniques developed for producing ultracold molecules,
photoassociation of ultracold atoms \cite{fio98,tak98,nik00,gab00,pic04} has 
proven its success in creating ultracold ($T \simeq 100~\mu$K) polar neutral 
molecules. Indeed, using that technique, magneto-optical trapping of ultracold polar 
neutral ground state KRb \cite{mancini04} and NaCs \cite{haim04} molecules, 
as well as formation of RbCs$^{\ast}$ molecules from a 
laser-cooled mixture of $^{85}$Rb and $^{133}$Cs atoms \cite{ker04,kerma04} 
were recently reported.
Exothermic chemical reactions and 
vibrational relaxation triggered by collisions are important factors limiting 
the lifetime of molecules created by photoassociation in highly excited vibrational 
levels \cite{bal01,bod02}. Although collisional studies of ultracold molecules 
have been a matter of active research in recent 
years \cite{balak98,balak00,balak03,stoe03,til04,sold02,volpi03}, relatively 
few progress has been reported on chemical reactivity of polar molecules at ultralow 
temperatures \cite{weck04,bala04}.  
   
In this work, we report quantum scattering calculations for the
$\mbox{Li}+\mbox{HF}\rightarrow\mbox{H}+\mbox{LiF}$ 
reaction at cold and ultracold translational energies.
Since methods for cooling and trapping alkali metal atoms have reached 
high degree of sophistication and creation of BEC of alkali metal atoms has 
become rather widespread, collisions of ultracold alkali metal atoms with polar 
molecules are being explored as a possible method for creating ultracold polar 
molecules. 
Thus, cross sections for elastic and ro-vibrationally inelastic collisions of 
$\mbox{Li}+\mbox{HF}$ system are of significant interest. Moreover, 
from a chemical dynamics point of view the $\mbox{Li}+\mbox{HF}$ collision is 
especially interesting due to the unusually deep van der Waals minimum of 
about $0.24~\mbox{eV}~(1936~\mbox{cm}^{-1})$ in the entrance channel of the collision. Since 
$\mbox{Li}+\mbox{HF}\rightarrow\mbox{LiF}+\mbox{H}$ involves the transfer of 
the relatively heavy F atom (the $\mbox{LiH}+\mbox{F}$ channel is highly endoergic 
and is not open at low energies), it will be particularly interesting to see 
whether the reaction will occur with significant rate coefficient at ultralow 
energies. 

The $\mbox{Li}+\mbox{HF}$ reaction has been the topic of a large number of 
experimental and theoretical studies.
After the pioneering crossed beam work of \citet{tayl55}, the $\mbox{Li}+\mbox{HF}$ 
reaction became a 
prototype system for experimental studies of the ``harpoon'' mechanism 
in reactions between alkali or alkaline earth metal atoms and hydrogen 
halide molecules \cite{hers66}. Thus, a large amount of experimental information 
has been reported for key observables such as integral and differential reactive cross 
sections \cite{beck80,loes93,baer94,aoiz99,aoiz00,casa00,hobe01,aoiz01,hobe04}.
On the theoretical front, numerous quantum mechanical 
\cite{baer94,park95,gogt96,agua97a,agua97b,lara98,pani98,aoiz99,aoiz00,lara00,
aoiz01,wei03,laga04} 
as well as classical trajectory \cite{aoiz00,aoiz00b} scattering 
calculations have been performed on the ground state potential 
energy surface (PES).
The relative simplicity of the LiHF system, with only 
13 electrons, makes it very suitable for accurate {\em ab initio} calculations. 
Consequently, a rich variety of analytic global fits to the $X^2A'$ symmetry 
electronic ground state PES have been proposed \cite{zeir78,chen80,cart80,laga84,garc84,
pani89,palm89,suar94,agua95,agua97a,agua97b,laga98a,laga98b,burc00,jasp01,jasp02,
burc02,
agua03}. 
As mentioned above, one of the unique aspects of the LiHF system is the rather 
deep van der Waals 
well in both the $\mbox{Li}+\mbox{HF}$ and $\mbox{H}+\mbox{LiF}$ channels. Unlike the 
well studied $\mbox{F}+\mbox{H}_2$ and $\mbox{Cl}+\mbox{H}_2$ systems where the van 
der Waals well depth is about $100-200~\mbox{cm}^{-1}$, the van der Waals well in 
the $\mbox{Li}+\mbox{HF}$ is an order of magnitude deeper, giving rise to long-lived 
collision complexes and narrow scattering resonances in the energy dependent reaction 
probabilities. 
The presence of the deep van der Waals well in the 
$\mbox{Li}(^2S)+\mbox{HF}(X ^1\Sigma^+)$ entrance valley was confirmed 
by backward glory scattering experiment of \citet{loes93} and by  
spectroscopic measurements of \citet{huds00}.

The $X^2A'$ LiHF PESs used in previous scattering 
studies were based on a relatively restricted sets of {\em ab initio} data, thus 
limiting the accuracy of the calculations. Furthermore, the energy range investigated did 
not cover the translationally cold and ultracold regimes. 
Here, we report quantum scattering calculations  
for $\mbox{Li}(^2S)+\mbox{HF}(X ^1\Sigma^+; v=0,1,j=0)\rightarrow\mbox{H}+
\mbox{LiF}(X ^1\Sigma^+; v',j')$ collisions, for a total molecular angular momentum $J=0$, 
using the recent 
high accuracy global PES of the LiHF ground state calculated by \citet{agua03}.  
A brief review of the basic characteristics of the PES is given in Sec. II, together  
with a summary of the quantum scattering approach with illustrative convergence 
tests assessing the validity of our calculations. In Sec. III, we present 
state-to-state and initial-state-selected probabilities, cross sections, and 
rate coefficients for both reactive and non-reactive open channels of the  
collision.     
We discuss the effect of vibrational excitation on chemical reactivity 
at low temperatures and provide a summary of our findings in Sec. IV.

\section{Calculations}  \label{sec:1}

\subsection{$X^2A'$ state potential energy surface}  

Calculations reported in the present study have been carried out using the 
recent LiHF ground state PES of \citet{agua03}. This chemically accurate PES 
was computed for about 6000 nuclear geometries using internally contracted 
multireference configuration interaction (MRCI) wave functions including all 
single and double excitations and Davidson size consistency correction (+Q).
A large atomic basis set was used to adequately describe the 
$\mbox{Li}^++\mbox{HF}^-$ and $\mbox{Li}^++\mbox{H}^-\mbox{F}$ ionic configurations 
responsible for the curve crossing leading to the LiF products in the 
adiabatic representation of the electronic ground state. A saddle point results 
from the crossing between the $\mbox{Li}^++\mbox{HF}^-$ ionic state and a covalent 
configuration correlating to $\mbox{Li}(^2S)+\mbox{HF}(X ^1\Sigma^+)$. On the 
basis of these MRCI+Q results, an analytic global PES was constructed using 
the modified many-body expansion of \citet{agua92}.
Major features of this PES are as follows: a $-0.241~\mbox{eV}$ deep 
van der Waals well corresponding to the Li$\cdots$FH complex in the 
entrance channel due to strong dipole electric fields of the reagents 
followed by a saddle point at $+0.251~\mbox{eV}$. 
The formation of the LiF$\cdots$H complex takes place in a late shallow van der 
Waals well with a minimum at $+0.118~\mbox{eV}$ in the product 
valley, connecting with the $\mbox{H}(^2S)+\mbox{LiF}(X ^1\Sigma^+)$ products 
asymptote at $+0.186~\mbox{eV}$. All energies are relative to the 
$\mbox{Li}(^2S)+\mbox{HF}$ asymptote with energy $E=0$ corresponding to the 
bottom of the HF potential.
Thus, the 
$\mbox{Li}+\mbox{HF}\rightarrow\mbox{H}+\mbox{LiF}$ reaction is endoergic with 
exclusion of the zero-point energy of the reactants and products. The reaction 
becomes exoergic with ground state reagents if the zero-point energy of the reactants 
and products is included. The exoergicity is $0.01122~\mbox{eV}$ with ground state 
reagents.   
The $\mbox{LiH}(X ^1\Sigma^+)+\mbox{F}(^2P)$ products 
lie at $3.57~\mbox{eV}$ and this reaction channel is 
closed for the energy range covered in this study.

\subsection{Quantum scattering calculations}  

Quantum reactive scattering calculations have been performed using the ABC
program developed by \citet{sko00}. This implementation of the coupled-channel 
hyperspherical coordinate method solves the Schr\"{o}dinger equation in 
Delves hyperspherical coordinates for the motion of the three nuclei on the 
parametric representation of a single Born-Oppenheimer PES with reactive 
scattering boundary conditions applied exactly. 

Parity-adapted $S-$matrix 
elements, $S^{J,P}_{v'j'k',vjk}$, are computed for all the arrangements of the 
collision products for each given $(J,P,p)$ triple, where $J$ is the total 
angular momentum quantum number and $P$ and $p$ are the triatomic and diatomic 
parity eigenvalues, respectively; $v$ and $j$ are the usual diatomic vibrational 
and rotational quantum numbers and $k$ is the helicity quantum number for 
the reactants, their primed counterparts referring to the products. After 
transformation of the parity-adapted $S-$matrix elements into their standard 
helicity representation, $S^{J}_{v'j'k',vjk}$, initial state selected cross sections 
are calculated as a function of the kinetic energy, $E_{kin}$, according to
\begin{equation}\label{stscs}
\sigma_{vj}(E_{kin}) = \frac{\pi}{k^2_{vj}(2j+1)}\sum^{J_{max}}_{J=0} 
(2J+1) \sum_{v'j'k'k} \vert S^{J}_{v'j'k',vjk}(E_{kin})\vert ^2,
\end{equation}
where $k_{vj}$ is the incident channel wave vector and the helicity quantum numbers 
$k$ and $k'$ are restricted to the ranges $0\leqslant k \leqslant \mbox{min}(J,j)$ and 
$0\leqslant k' \leqslant \mbox{min}(J,j')$. Let us note that for zero total molecular 
angular momentum and $s-$wave scattering in the incident channel, Eq. (\ref{stscs}) merely 
reduces to a summation over the quantum number $v'$ and $j'$.

\subsection{Convergence tests}  

At very low temperatures, quantum tunneling becomes the dominant mechanism  
of chemical reaction when energy barriers are present. As a consequence, the reaction 
probabilities are usually very small and particular care must be paid to the 
convergence of scattering calculations. 
We have performed extensive convergence tests of the initial-state-selected and 
state-to-state reaction probabilities with respect to the maximum rotational 
quantum number, $j_{max}$, and cut-off energy, $E_{max}$, that control the 
basis set size, the maximum value of the hyperradius, $\rho_{max}$, and 
the size of the log derivative propagation sectors, $\Delta{\rho}$.
    
The energy dependence of the 
$\mbox{Li}+\mbox{HF}(v=0,j=0)\rightarrow\mbox{H}+\mbox{LiF}(v',j')$
reaction probability is shown in Fig. \ref{fig1} for different values of 
$\rho_{max}$ and $\Delta{\rho}$. Convergence with an accuracy better than 
$10^{-10}$ was achieved over the range $10^{-5}-10^{-3}~\mbox{eV}$ using 
the values $\rho_{max}=50.0~\mbox{a.u.}$ and $\Delta{\rho}=0.005~\mbox{a.u.}$  
A more stringent convergence test consisted in the analysis of the product 
rotational distribution represented in Fig. \ref{fig2}. The same 
values of $\rho_{max}$ and $\Delta{\rho}$ as above were used to calculate 
the state-to-state reaction probabilities for   
$\mbox{Li}+\mbox{HF}(v=0,j=0)\rightarrow\mbox{H}+\mbox{LiF}(v'=0,j')$ 
at a fixed incident kinetic energy of $10^{-5}~\mbox{eV}$. 
Using the results obtained with $j_{max}=25$ and $E_{max}=3.2~\mbox{eV}$ as 
a reference, similar accuracy was found using $j_{max}=20$ and 
a cut-off internal energy $E_{max}=2.9~\mbox{eV}$ in any channel. 
The basis set corresponding to these values was composed of 771 local basis functions. 
As Fig. \ref{fig2} illustrates, the 
state-to-state reactive probability is particularly sensitive to the size 
of the basis set at low translational energies. 
On the basis of these convergence tests, values of $\rho_{max}=50.0~\mbox{a.u.}$, 
$\Delta{\rho}=0.005~\mbox{a.u.}$, $j_{max}=20$ and $E_{max}=2.9~\mbox{eV}$ 
were adopted for the calculations reported hereafter.

\section{Results and discussion}  \label{sec:2}

The initial state-selected reaction probability for LiF formation in 
$\mbox{Li}+\mbox{HF}(v=0,j=0)$ collisions is shown in Fig. \ref{fig3} 
as a function of the total energy. 
Our results are presented along with the recent time-independent quantum 
coupled channel hyperspherical calculations of \citet*{laga04} obtained 
with a scaled PES of \citet{park95}. Both sets of results are consistent 
with respect to the magnitude of the predicted probability, i.e., both exhibit 
small values for the reaction probability. This merely reflects the fact that for 
collisions with HF molecules initially in their ground vibrational state     
the reaction proceeds mainly by quantum tunneling through the barrier. 
The unusually large well depth of the van der Waals potential in the 
entrance valley effectively raises the reaction barrier, thus leading to small 
values of the reaction probability.  
Our results confirm that there is indeed a dense resonance structure at low 
energies associated with quasibound states of the $\mbox{Li}\cdots\mbox{F}-\mbox{H}$ 
van der Waals 
complex \cite{wei03,laga04}. However, the positions of the peaks predicted by our 
calculations are noticeably different from the quantum scattering results 
of \citeauthor{laga04} As discussed by \citet{agua97b}, the PES 
of \citeauthor{park95} used in most of the LiFH dynamical calculations 
performed until 1997, is based on a limited set of {\em ab initio} data 
and {\em ad hoc} modifications introduced to reproduce experimental 
properties resulted in artificial features in the PES. In addition, the 
total energy threshold for the time-independent calculations of \citet{laga04} 
is higher than our value of $0.2535~\mbox{eV}$ corresponding to the energy of the 
$\mbox{HF}(v=0,j=0)$ state. However, their time-dependent calculation carried out 
using wavepacket methods is 
in line with our prediction of the threshold position \citep[see][Fig. 3]{laga04}. 
Comparison of our results presented in Fig. \ref{fig3} with the recent quantum 
mechanical scattering calculations of \citet*{wei03} is also very revealing of the 
quantitative discrepancies introduced by the PES in dynamical studies at low 
temperatures. Their time-independent quantum calculations using the 
variational method employed the global {\em ab initio} PES of \citet{jasp01}.  
Briefly, this ground-state PES is characterized by a reactant van der Waals well 
at $-0.21~\mbox{eV}$ relative to the $\mbox{Li}(^2S)+\mbox{HF}$ 
asymptote, followed by a saddle point at $+0.35~\mbox{eV}$, a product van der Waals well 
at $+0.167~\mbox{eV}$ and finally a product asymptote 
at $+0.21~\mbox{eV}$. This potential also exhibits a second saddle point in the 
product valley at $+0.224~\mbox{eV}$.   
Compared to our results, the reaction probability obtained by \citeauthor{wei03} 
for LiF formation is smaller by more than an order of magnitude. 
This reflects the effect of a $0.1~\mbox{eV}$ higher barrier in the 
reactant channel as well as the presence of a second saddle point in the product valley 
of the PES of \citeauthor{jasp01} 

Fig. \ref{fig4}~shows the state-to-state reaction probabilities for 
$\mbox{LiF}(v',j')$ formation as a function of the product rotational 
quantum number, $j'$, in $\mbox{Li}+\mbox{HF}(v=1,j=0)$ collisions.
For a fixed incident kinetic energy of $10^{-5}~\mbox{eV}$, 
5 vibrational levels are energetically accessible in the diatomic products of 
the reactions, each of these levels supporting 20 rotational states as restricted 
by our cut-off value for $j_{max}$. 
The probability for $\mbox{LiF}$ formation is larger for intermediate-$j'$ product 
channels of the $v'=0$ and $v'=1$ vibrational levels.    
A broad peak centered at $j'=10$ appears in the population distribution of these 
vibrational states, corresponding to an exoergicity of 
$4.85\times 10^{-1}~\mbox{eV}=11.183~\mbox{kcal/mol}$ and 
$3.74\times 10^{-1}~\mbox{eV}=8.629~\mbox{kcal/mol}$  
for the reaction to $v'=0$ and $v'=1$, respectively. 
Vibrational excitation of the reactants significantly increases the reaction probability, 
as can be seen from the comparison of Fig. \ref{fig2} and Fig. \ref{fig4}. 

Initial-state-selected cross sections for LiF formation and for 
nonreactive scattering in $\mbox{Li}+\mbox{HF}(v=0,1,j=0)$ collisions are displayed 
in Fig. \ref{fig5} for incident translational energies covering the range 
$10^{-7}-10^{-1}~\mbox{eV}$. 
The reaction cross section is rather small for 
HF molecules initially in their ground vibrational state since quantum tunneling 
of the relatively heavy fluorine atom is the 
dominant reaction mechanism. For energies below $10^{-5}~\mbox{eV}$, 
the reaction cross section reaches the Wigner regime \cite{wig48} where it varies 
inversely as the velocity. 
However, a strong peak centered at $5\times 10^{-4}~\mbox{eV}$ is observed where 
the cross section increases by about six orders of magnitude. 
This feature suggests that reactivity may be more important at low temperatures 
than generally recognized. 
For translational energies beyond 
$10^{-3}~\mbox{eV}$, the reaction cross section is characterized by resonant spikes 
due to metastable states of the $\mbox{Li}\cdots\mbox{F}-\mbox{H}$ van der Waals 
complex in the initial 
channel. Nonreactive channels of the $\mbox{Li}+\mbox{HF}(v=0,j=0)$ collisions 
are open only for translational energies larger than $5.06\times 10^{-3}~\mbox{eV}$, 
thereby explaining the sharp rise in the nonreactive cross section at this value 
corresponding to the energy for rotational excitation to the first excited state of the 
product, $\mbox{HF}(v=0,j=1)$. Beyond this energy threshold, nonreactive scattering 
becomes more favorable than LiF formation, as shown in the lower panel of 
Fig. \ref{fig5}. On the contrary, in $\mbox{Li}+\mbox{HF}(v=1,j=0)$ collisions 
the reactive channel dominates the nonreactive processes, with a $\mbox{LiF}/\mbox{HF}$ 
product branching ratio reaching 20 at low and ultralow temperatures. 
This is especially interesting as the reaction involves quantum tunneling of the 
relatively heavy fluorine atom.
Moreover, chemical reactivity is greatly enhanced by vibrational excitation.
In the Wigner regime, where cross section ratios become constant, the reaction cross 
section involving excited $\mbox{HF}(v=1,j=0)$ reactants are 635 times larger than 
for collisions with HF reactants in their rovibrational ground state.

Further characterization of the peaks represented in the lower panel of 
Fig. \ref{fig5} has been carried out
by calculating the bound- and quasi-bound states of the 
$\mbox{Li}\cdots\mbox{F}-\mbox{H}$ van 
der Waals potential that correlate with the $\mbox{HF}(v=0)$ manifold.
The adiabatic potentials are obtained by constructing the matrix elements of the 
interaction potential in a basis set of the rovibrational levels of the HF molecule 
and diagonalizing the resulting diabatic potentials as a function of the 
atom-molecule separation, $R$. The resonance energies and the corresponding 
wave functions are computed using the Fourier grid Hamiltonian 
method \cite{mars89,bali92}.   
For constructing the adiabatic potentials, we used a 20-term Legendre expansion 
of the interaction 
potential, 25 angular orientations to project out the expansion coefficients, 
17 Gauss-Hermite 
quadrature points for the vibrational wave functions and a grid of 1000 points 
for the atom-molecule separation.
As reported in Table \ref{tab1} and Fig. \ref{fig6}, 
the excellent agreement found between the energy eigenvalues and the peak positions 
from our scattering calculations suggests that peaks A to H in Fig. \ref{fig6} are 
resonances due to the decay of metastable states of the 
$\mbox{Li}\cdots\mbox{HF}$ van der Waals complex. 
The resonances correspond to quasibound states of the adiabatic potentials correlating 
with $j=1-4$ of the $\mbox{Li}\cdots\mbox{HF}(v=0)$ molecule. Each of the adiabatic potential supports 
a number of quasibound  complexes due to the relatively deep van der Waals interaction 
in the entrance channel.
Only high-lying stretching 
vibrational states of the van der Waals complex generate resonances for $j=1$ 
$(t=10,11)$ and $j=2$ $(t=5,6,8)$, while low$-t$ channels 
give rise to resonances for $j=3,4$.
However, we have not been able to assign the strong peak centered at       
$E=2.539\times 10^{-1}~\mbox{eV}$ to a reactant van der Waals complex. 
Nevertheless, time-delay calculations show that it is a reactive scattering resonance. 

Elastic cross sections for $s-$wave scattering 
in $\mbox{Li}+\mbox{HF}(v=0,1,j=0)$ collisions are presented  
in Fig. \ref{fig7} as a function of the incident translational energy. 
For translational energies below $10^{-3}~\mbox{eV}$ elastic cross sections 
for $v=0$ and 1 are nearly identical. Above this energy value, 
the results for $v=1$ are less oscillatory compared to those of $v=0$. The real 
part of the scattering length has been calculated for $v=0$ and 1 in the 
ultracold limit according to
\begin{equation}\label{rscatl}
\alpha = -\lim_{k\to 0}\frac{\mbox{Im}(S^{el})}{2k},
\end{equation}
where $S^{el}$ is the elastic component of the scattering matrix and $k$ is 
the wavevector corresponding to the initial kinetic energy. We found 
$\alpha_{v=0}=+11.551~\mbox{\AA}$ and $\alpha_{v=1}=+11.535~\mbox{\AA}$ for the real part of 
the scattering length for $v=0$ and 1, respectively.       

Fig. \ref{fig8}~shows the $J=0$ contribution to the reaction rate coefficients 
for $\mbox{LiF}(v',j')$ formation in $\mbox{Li}+\mbox{HF}(v=0,1,j=0)$ 
collisions, evaluated as the product of the cross section and the relative velocity, 
as a function of the translational temperature, $T=E_{kin}/k_B$, where $k_B$ is 
the Boltzmann constant. 
The unusually large well depth of the van der Waals potential in the 
entrance channel ($-0.2407~\mbox{eV}$, relative to the $\mbox{Li}(^2S)+\mbox{HF}$ 
asymptote of the PES) effectively raises the reacton barrier and lead to small 
values of the reaction rate coefficients at low energies. 
For $\mbox{HF}(v=0,j=0)$ reactants, the rate coefficient reaches the Wigner 
regime for temperatures below $0.03~\mbox{K}$, with a constant value of 
$4.5\times 10^{-20}~\mbox{cm}^3~\mbox{s}^{-1}$ in the zero-temperature limit. 
Vibrational excitation to the $v=1$ state enhances reactivity by 3 orders of 
magnitude in the ultracold limit, 
as depicted in the upper panel of Fig. \ref{fig8}, where a 
constant value of $2.8\times 10^{-17}~\mbox{cm}^3~\mbox{s}^{-1}$ is attained for 
the reaction rate coefficient for temperatures below $0.005~\mbox{K}$. For both $v=0$ and 1, 
the reactivity rapidly increases beyond $1~\mbox{K}$. 
However, accurate prediction of rate coefficients for higher temperatures requires 
calculations for $J>0$ which is beyond the scope of this work.

\section{Conclusion}  \label{sec:conclu}

Quantum reactive scattering calculations have been performed for the 
$\mbox{Li}+\mbox{HF}(v=0,1,j=0)\rightarrow\mbox{H}+\mbox{LiF}(v',j')$ 
bimolecular scattering process for zero total angular momentum, at low and 
ultralow temperatures. 
The energy dependence of state-to-state and initial-state-selected 
probabilities and cross sections, as well as limiting values of the rate 
coefficients have been evaluated using the 
coupled-channel hyperspherical coordinate method. 
For $\mbox{Li}+\mbox{HF}(v=0,j=0)$ collisions, our calculations, 
using the most recent PES for the LiHF electronic ground state, clearly illustrate  
the dominance of the resonance tunneling mechanism due to the decay of metastable 
states of the Li$\cdots$HF van der Waals complex in the entrance valley into the 
$\mbox{LiF}(v'=0)$ product manifold. 
Comparison of our calculations with previous quantum scattering results 
emphasizes the extreme sensitivity of scattering matrix elements to the 
details of the PES and therefore the desirability for high accuracy analytic 
fits to correctly describe the collision dynamics and the interplay among the 
various energy modes in the cold and ultracold regimes. 
We also find that chemical reactivity 
is dramatically enhanced by vibrational excitation for cold and ultracold 
translational energies, with a 3-order of magnitude increase between 
the $v=0$ and $v=1$ rate coefficients in the zero-temperature limit, 
consistent with our findings for the $\mbox{H}+\mbox{HCl}$ and 
$\mbox{H}+\mbox{DCl}$ reactions \cite{weck04}.   
Moreover, our results show that the LiF formation dominates the nonreactive 
processes in $\mbox{Li}+\mbox{HF}(v=1,j=0)$ collisions, with a 
$\mbox{LiF}/\mbox{HF}$ product branching ratio reaching 20 at low and 
ultralow temperatures.

The rich resonance features characterizing the energy dependence of the 
$\mbox{Li}+\mbox{HF}$ cross sections make the LiHF system particularly 
attractive for the study of coherent control of resonance-mediated reactions. 
In fact, recent advances in the control of bimolecular processes have 
shown that cross sections resulting from scattering that proceeds via an intermediate 
resonance are exceptionally controllable \cite{zema04}. This offers new 
possibility for tuning chemical reactivity at the single quantum state level 
of resolution. The present study shows that vibrational excitation may be used to 
circumvent reaction barriers at cold and ultracold temperatures even when 
the reaction involves tunneling of a heavy atom such as fluorine.

\begin{acknowledgments}

This work was supported by NSF grant PHYS-0245019, the Research Corporation and by the 
United States-Israel Binational Science Foundation.

\end{acknowledgments}



\clearpage


\begin{table}
\caption{\label{tab1} Assignment of the resonances in the total energy dependence of the 
cross sections for LiF formation in $\mbox{Li}+\mbox{HF}(v=0,j=0)$ collisions (energies in eV).}
\begin{ruledtabular}
\begin{tabular}{cccccc}
\multicolumn{1}{c}{} &\multicolumn{1}{c}{} &\multicolumn{1}{c}{} 
& \multicolumn{3}{c}{Quantum numbers} \\ 
\cline{4-6} \\ 
 \multicolumn{1}{c}{Resonance} 
& \multicolumn{1}{c}{Peak} 
& \multicolumn{1}{c}{Binding energy of} 
& \multicolumn{1}{c}{$v$\footnotemark[2]} & 
\multicolumn{1}{c}{$j$\footnotemark[3]} & 
\multicolumn{1}{c}{$t$\footnotemark[4]}
\\
 \multicolumn{1}{c}{} 
& \multicolumn{1}{c}{position} 
& \multicolumn{1}{c}{$\mbox{Li}\cdots\mbox{HF}(v,j)$ complex\footnotemark[1]} 
& \multicolumn{1}{c}{} & 
\multicolumn{1}{c}{} & 
\multicolumn{1}{c}{}
\\

\hline
 A & 0.2549& 0.2549 & 0 & 2 & 5\\
 B & 0.2554& 0.2553 & 0 & 1 & 10\\
 C & 0.2568 & 0.2568 & 0 & 3 & 2\\
 D & 0.2579 & 0.2578 & 0 & 1 & 11\\
 E & 0.2587 & 0.2585 & 0 & 4 & 0\\ 
 F & 0.2597 & 0.2596 & 0 & 2 & 6\\
 G & 0.2656 & 0.2646 & 0 & 3 & 3\\
 H & 0.2665 & 0.2664 & 0 & 2 & 8\\
\end{tabular}
\end{ruledtabular}
\footnotetext[1]{Energies are calculated with the Fourier grid Hamiltonian method. 
 Energies are relative to separated $\mbox{Li}+\mbox{HF}$ system with energy zero 
 corresponding to the bottom of the HF potential.}
\footnotetext[2]{HF vibrational quantum number.}
\footnotetext[3]{HF rotational quantum number.}
\footnotetext[4]{Quantum number for the $\mbox{Li}-\mbox{HF}(v,j)$ van der Waals stretching 
vibration.}
\end{table}

\clearpage


\begin{figure} 
\caption{\label{fig1} Translational energy dependence of the 
 reaction probability for $\mbox{LiF}(v',j')$ formation in $\mbox{Li}+\mbox{HF}(v=0,j=0)$ 
 collisions for different values of $\rho_{max}$ and $\Delta{\rho}$. }
\end{figure}

\begin{figure} 
\caption{\label{fig2} State-to-state reaction probability for 
 $\mbox{LiF}(v'=0,j')$ formation as a function of the product rotational 
 quantum number, $j'$, in $\mbox{Li}+\mbox{HF}(v=0,j=0)$ collisions. 
 Results are presented for various values of $E_{max}$ and $j_{max}$, at a
 fixed incident kinetic energy of $10^{-5}~\mbox{eV}$, $\rho_{max}=50.0~\mbox{a.u.}$, 
 and $\Delta{\rho}=0.005~\mbox{a.u.}$ }
\end{figure}

\begin{figure} 
 \caption{\label{fig3} Initial state-selected reaction probability for 
 LiF formation in $\mbox{Li}+\mbox{HF}(v=0,j=0)$ collisions 
 as a function of the total energy. Solid curve: present calculations; 
 dotted curve: quantum scattering calculation of \citet{laga04} (extracted 
 graphically from Fig. 3 of ref. \cite{laga04}).}
\end{figure}

\begin{figure} 
\caption{\label{fig4} State-to-state reaction probabilities for 
 $\mbox{LiF}(v',j')$ formation in $\mbox{Li}+\mbox{HF}(v=1,j=0)$  
 collisions as a function of the 
 product rotational number $j'$ for a fixed incident kinetic energy of 
 $10^{-5}~\mbox{eV}$. }
\end{figure}

\begin{figure} 
\caption{\label{fig5} Cross sections for LiF formation and nonreactive 
 scattering in $\mbox{Li}+\mbox{HF}(v,j=0)$ collisions, for $v=0$ (lower panel) 
 and $v=1$ (upper panel), as a function of the incident kinetic energy. 
 Dashed curve: nonreactive scattering; solid curve: LiF product channel.}
\end{figure}

\begin{figure} 
\caption{\label{fig6} Adiabatic potential energy curves and corresponding 
 quasibound levels of the $\mbox{Li}\cdots\mbox{HF}(v=0,j)$ van der Waals complex 
 (left panel); cross section for LiF formation 
 in $\mbox{Li}+\mbox{HF}(v=0,j=0)$ collisions as a function 
 of the total energy (right panel). The resonances A to H in the cross section 
 appear as a result of the decay of quasibound states of the Li$\cdots$HF van der 
 Waals complex.}
\end{figure}

\begin{figure} 
\caption{\label{fig7} Elastic cross sections for $s-$wave scattering in  
 $\mbox{Li}+\mbox{HF}(v=0,1,j=0)$ collisions as a function of the 
 incident kinetic energy. Solid curve: $v=0$; dashed curve: $v=1$.}
\end{figure}

\begin{figure} 
\caption{\label{fig8} Temperature dependence of reaction rate 
 coefficients for LiF formation in 
 $\mbox{Li}+\mbox{HF}(v=0,1,j=0)$ collisions. }
\end{figure}


\clearpage

\begin{figure*}
\begin{center}
\includegraphics[height=14cm,width=16cm]{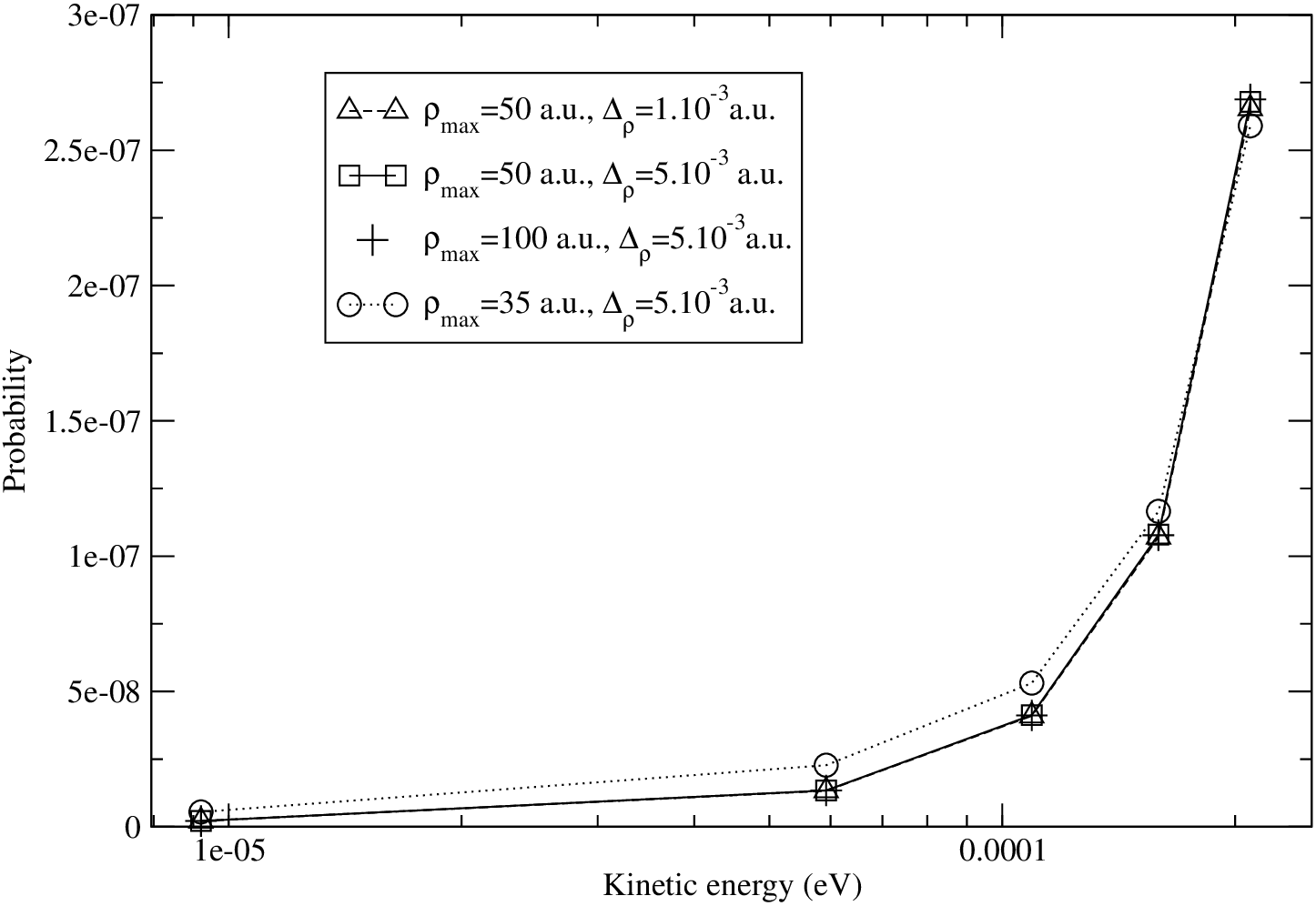}
\end{center} 
\end{figure*}

\clearpage
\begin{figure*}
\begin{center}
\includegraphics[height=14cm,width=16cm]{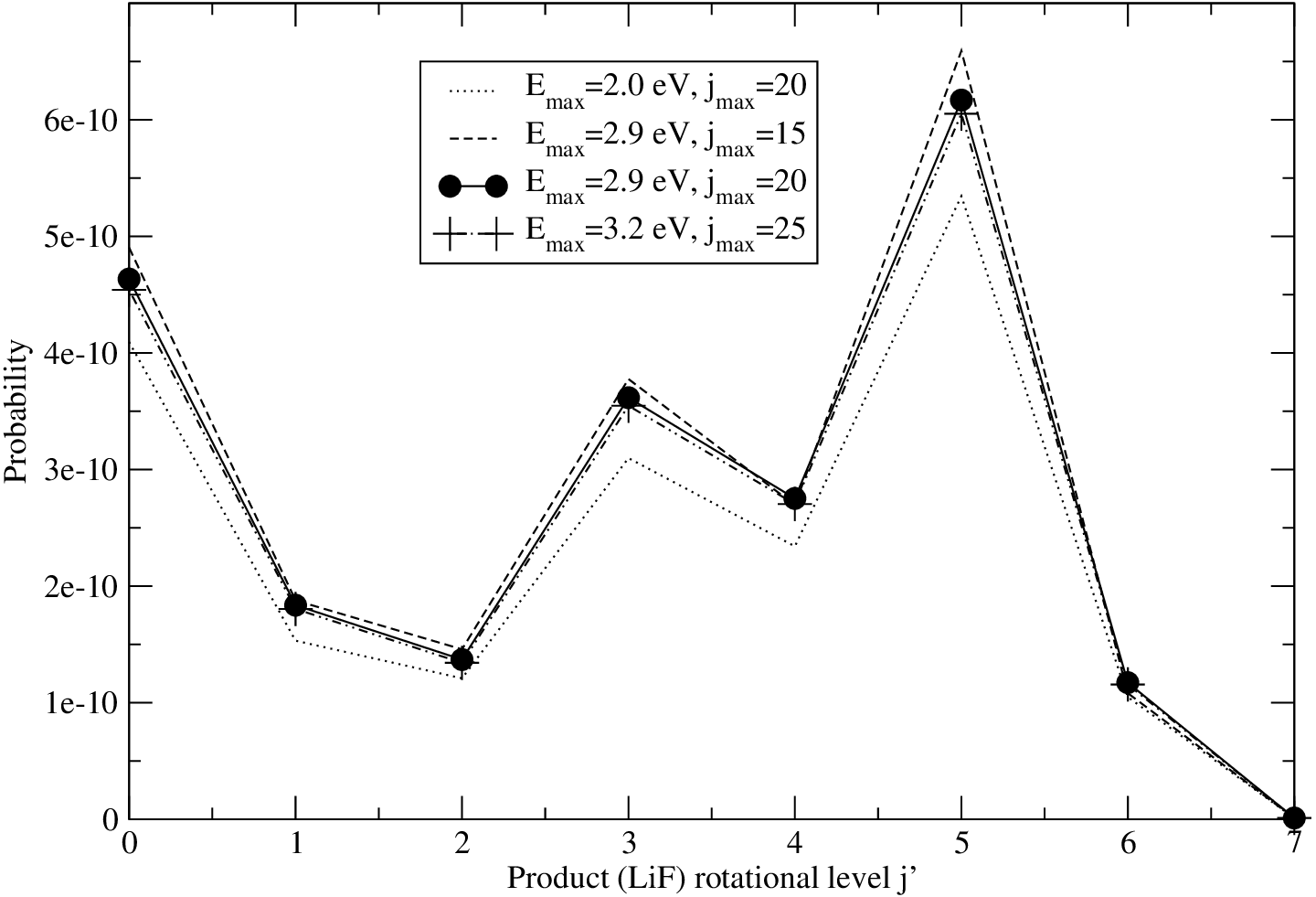}
\end{center} 
\end{figure*}

\clearpage
\begin{figure*}
\begin{center}
\includegraphics[height=14cm,width=16cm]{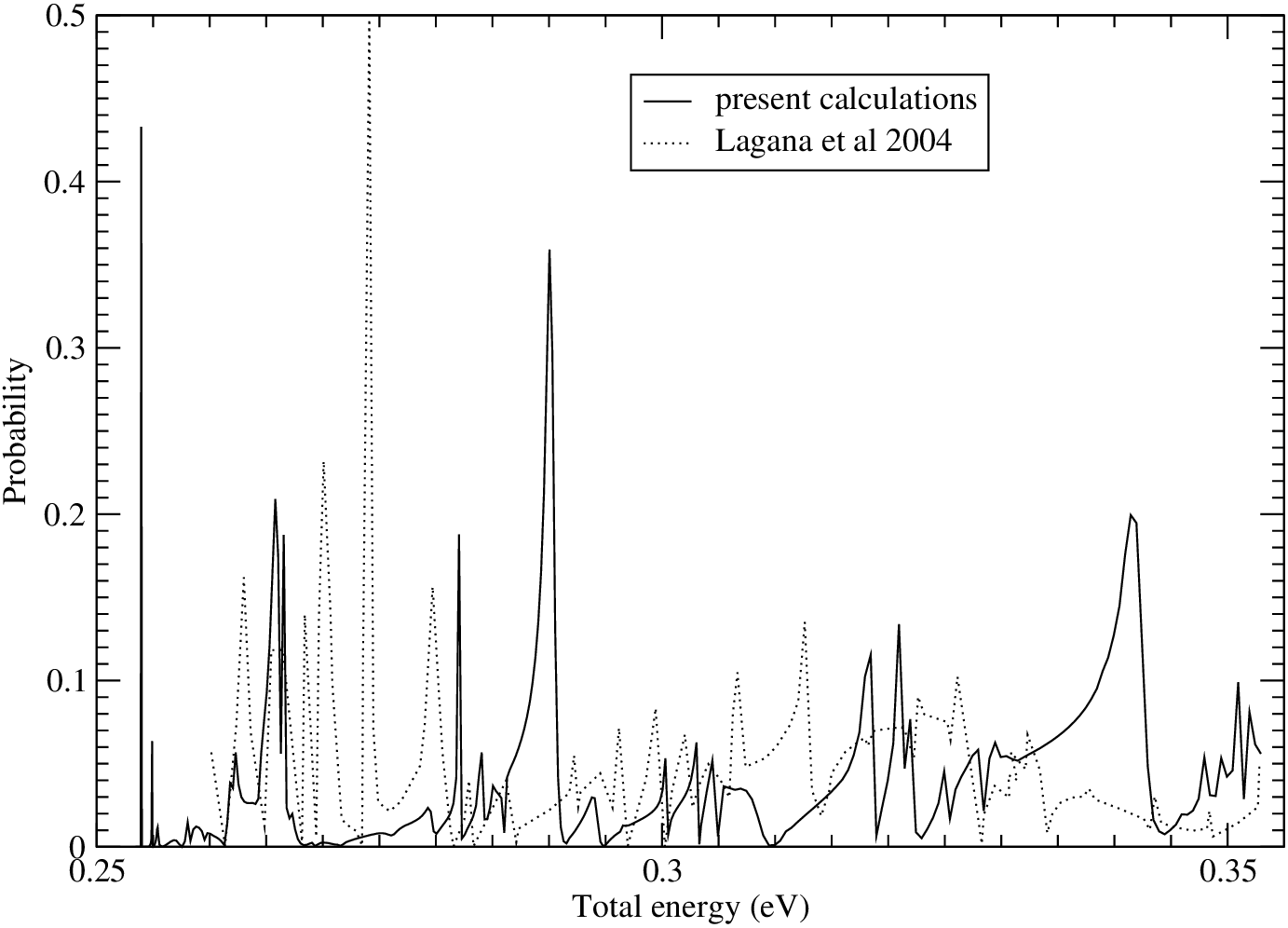}
\end{center} 
\end{figure*}

\clearpage
\begin{figure*}
\begin{center}
\includegraphics[height=14cm,width=16cm]{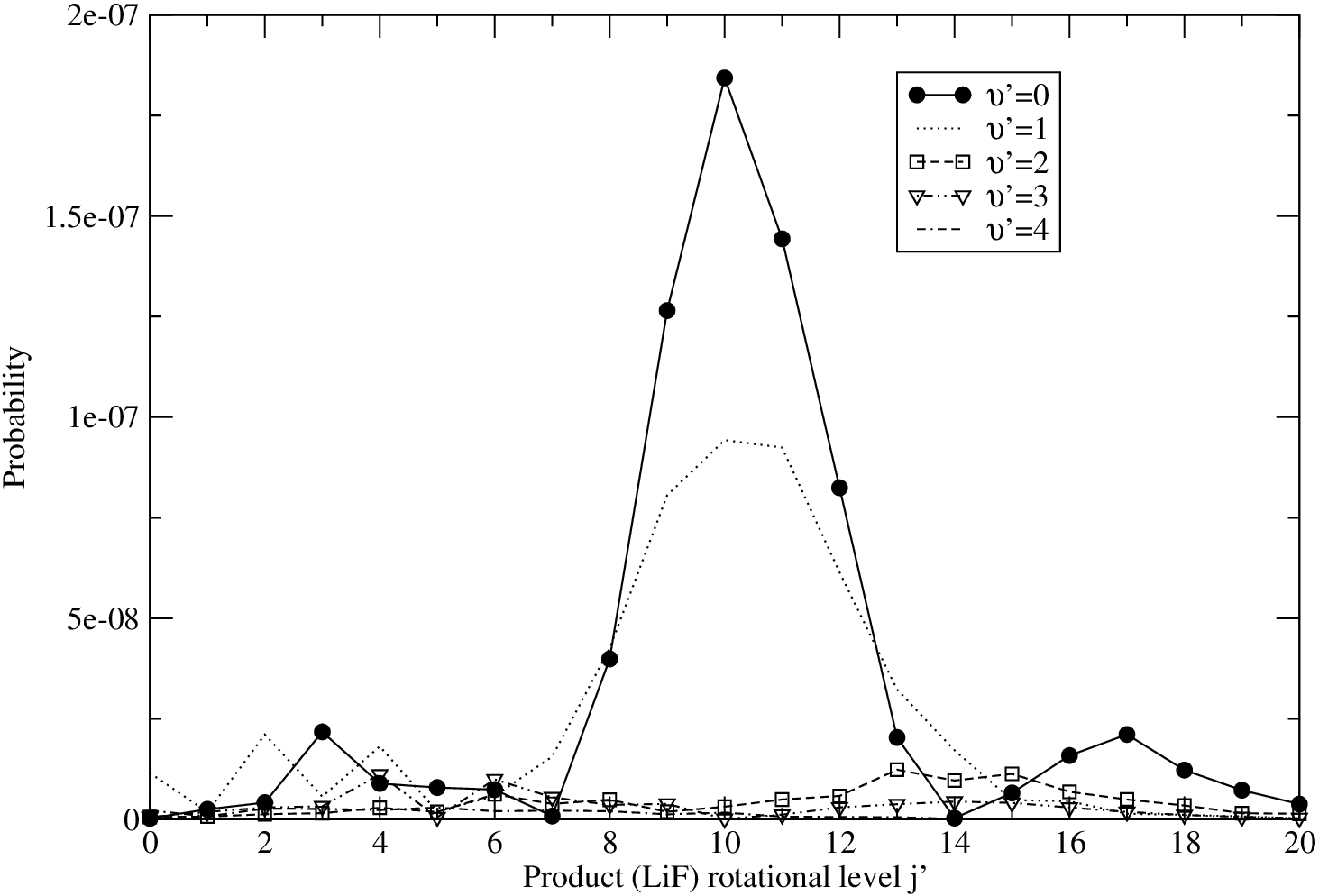}
\end{center} 
\end{figure*}

\clearpage
\begin{figure*}
\begin{center}
\includegraphics[height=14cm,width=16cm]{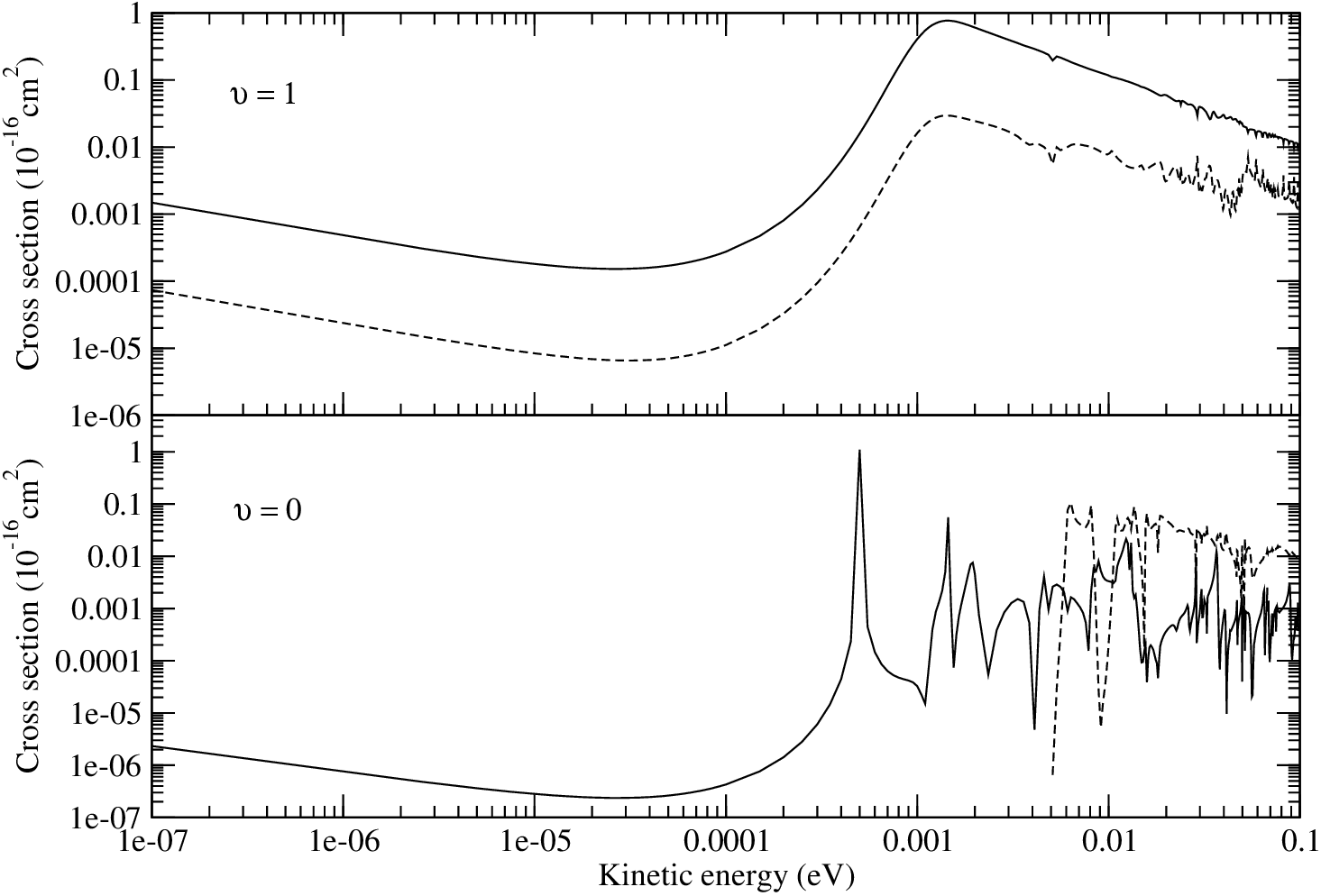}
\end{center} 
\end{figure*}

\clearpage
\begin{figure*}
\begin{center}
\includegraphics[height=14cm,width=16cm]{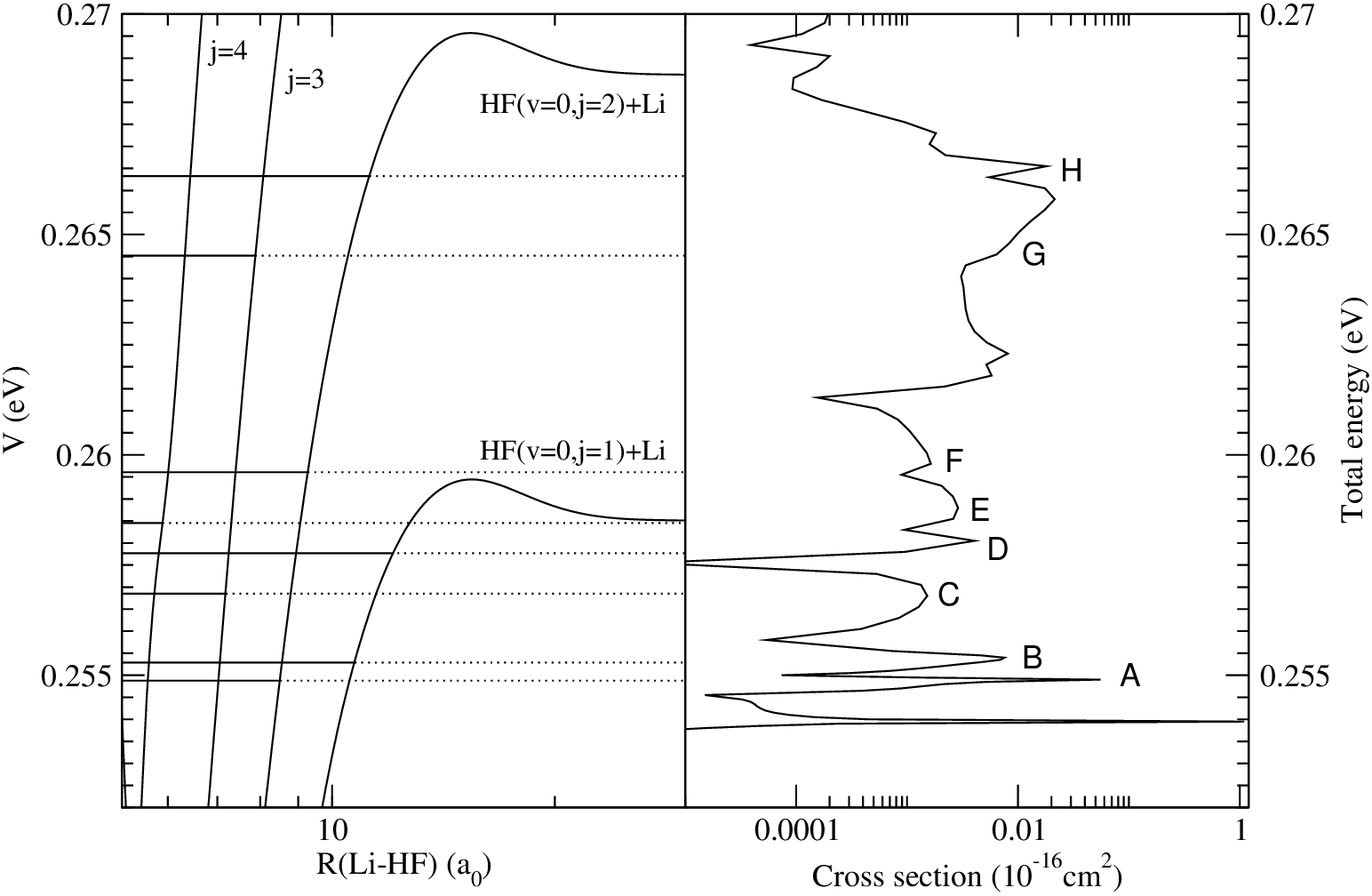}
\end{center} 
\end{figure*}

\clearpage
\begin{figure*}
\begin{center}
\includegraphics[height=14cm,width=16cm]{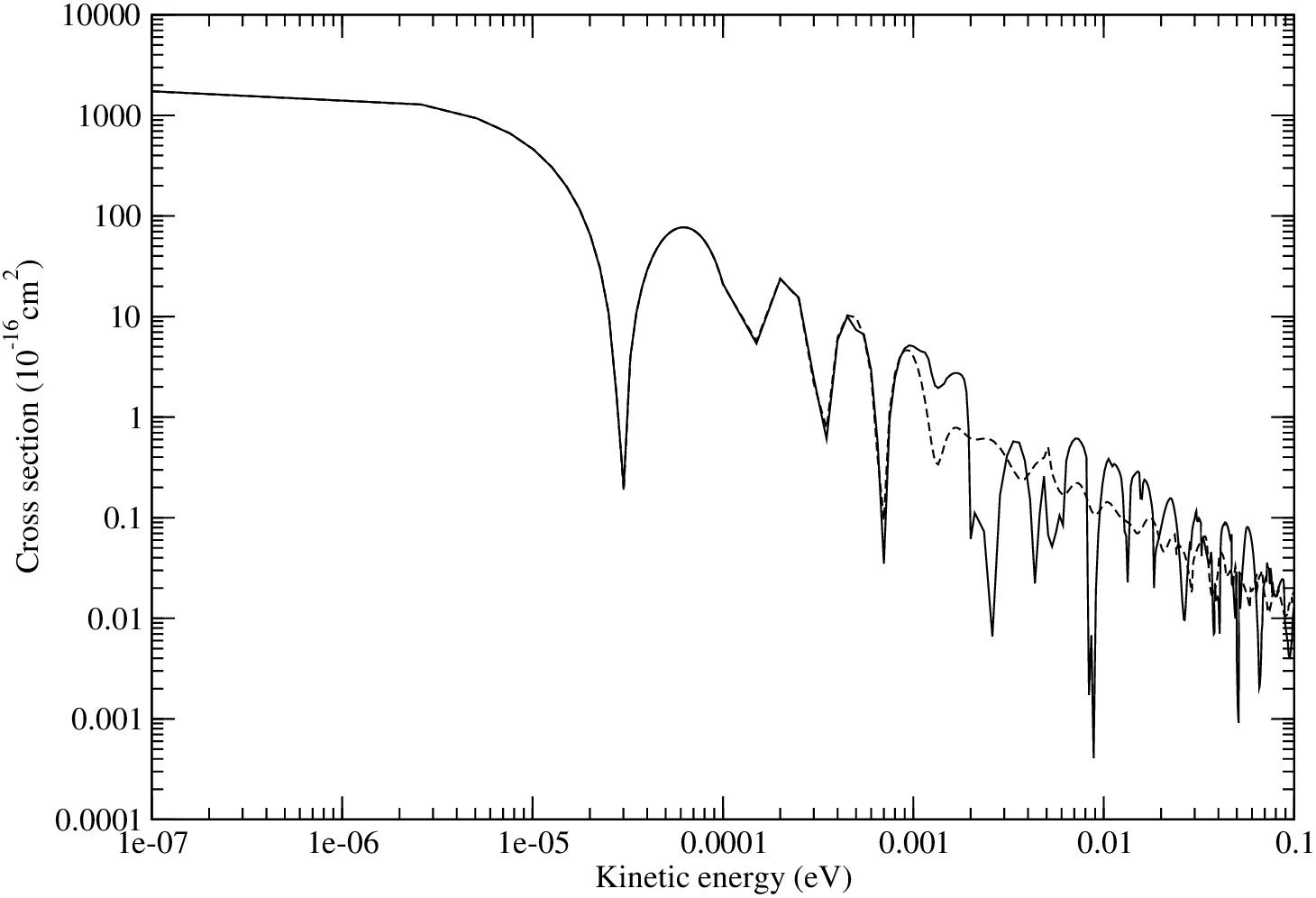}
\end{center} 
\end{figure*}

\clearpage
\begin{figure*}
\begin{center}
\includegraphics[height=14cm,width=16cm]{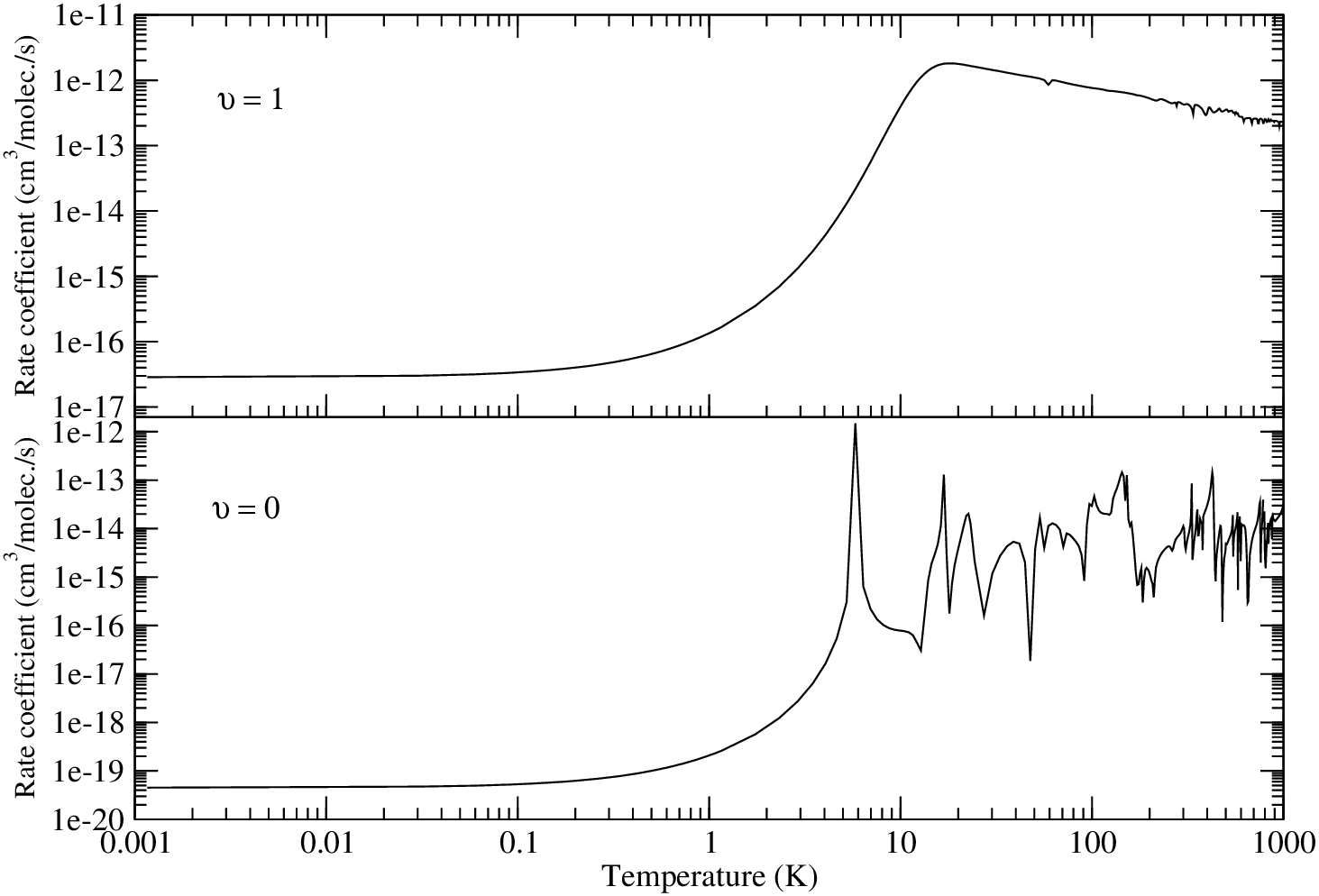}
\end{center} 
\end{figure*}

\end{document}